\documentstyle[12pt]{article}
\newcommand{\eqnsection}{
\renewcommand{\theequation}{\thesection.\arabic{equation}}
\makeatletter
\csname $addtoreset\endcsname
\makeatother}
\def\theequation{\thesection.\arabic{equation}}

\def\ll {\label}
\def\re{\ref}
\def\c{\cite}

\def\r1{(\ref{$1})}

\def\ti{\tilde}

\def\th{\theta}
\def\ba{\begin{array}{c}}

\def\sk{\smallskip}
\def\ea{\end{array}}

\def\ni{\noindent}

\def\bet{\beta}
\def\ov{\over}
\def\ha{{1\over 2}}

\def\l{\left}
\def\l({\left(}
\def\r){\right)}
\def\r{\right}
\def\rw{\rightarrow}

\def\la{\lambda}
\def\al{\alpha}

\def\be{\begin{equation}}
\def\bc{\begin{center}}
\def\ec{\end{center}}
\def\bit{\begin{itemize}}
\def\eit{\end{itemize}}
\def\ee{\end{equation}}
\def\ed{\end{document}}
\def\bea{\begin{eqnarray}}
\def\eea{\end{eqnarray}}
\def\d{\downarrow}
\def\up{\uparrow}
\begin{document}
\title{Integrability and exact solution of  correlated hopping
multi-chain electron systems} 
\author { Anjan Kundu\footnote {e-mail: anjan@tnp.saha.ernet.in}\footnote
{Address during Sept.-Dec., 2001 : Inst. Theor. Phys.,
 Uni-Dortmund. Germany} \\
Saha Institute of Nuclear Physics, 1/AF Bidhan Nagar,
\\700 064  Calcutta, India.}
\maketitle
\begin{abstract}

Exact quantum integrability  is established for a class of multi-chain
 electron models with
  correlated hopping and spin models with interchain interactions, by
  constructing the related Lax operators and
$R$-matrices
through twisting and gauge
 transformations.  Exact solution of the eigenvalue problem for commuting
conserved
 quantities of such systems is achieved through algebraic Bethe ansatz, on
the examples of Hubbard and $t-J$ models with correlated hopping.
 Our systematic construction identifies the integrable subclass of such
 known solvable models and also generates new systems including the
generalized $t-J$ models. At the same time it makes proper correction to a
well known model and resolves recent controversies regarding the equivalence
and solvability of some known models.
 \end{abstract}
\ni {\it PACS}: 02.30.IK,
04.20.Jb,
  71.10.Pm,
05.30.Pr
\sk

\ni {\it Keywords}: Quantum integrable systems, Algebraic Bethe ansatz,
Gauge and twisting transformations, Correlated hopping electron and spin
models
\section{Introduction}
\setcounter{equation}{0}
There is an  upsurge of interest in recent years in the study of
correlated electron and spin systems in low dimensions, which is  
 motivated mostly by  the recent possibility of their fabrication and
experimental verification of related theoretical results \c{ldsystems},
 as well as by their potential for 
applications
 to the  high $T_c$-superconductivity
\c{hald81}.   
On the other hand one can take an exclusive advantage   in one-dimension,
since  some class of models in this case may become
   exactly solvable and even completely integrable, which therefore 
 can enhance our understanding of the low-dimensional physics
 by  providing  detailed picture 
of the system through  exact
results. 
Few well known examples of such models are the $XXX$ and $XXZ$ spin chains
\c{schains}, 
  the Hubbard model \c{hubbard,hubbardi}, the $t-J$
model \c{sarkar,koress} etc.,
along with their various  extensions and the ladder or their multi-chain
generalisations \c{korepin,spinlad,tjlad}.
Recently, a large class of  models describing electrons with correlated
hopping, spin ladders with interchain interactions and their multi-chain
generalizations have attracted special attention due to their simplicity and 
 exact solvability
\c{borov912,zvyag92,SS98,kun98,oster00}. A 
  systematic analysis has also been carried out   \c{oster00} 
  to identify the general  class of these models which  exhibits 
  exact solvability by coordinate Bethe ansatz (CBA).
  However, the
 important question of complete quantum integrability for such systems,
 which defines a much richer class of models allowing infinite set of
 commuting conserved operators along with their   exact eigenvalue
 solution through 
   the algebraic Bethe ansatz (ABA) \c{qis}, has not been
explored. Moreover, the recent controversy around the solvability
\c {oster00} as well as the  claim and
denial  of the equivalence between some models \c{zvyag99,SS99}
  could not be  solved satisfactorily  
due to the lack of understanding of
the general structures for such systems.

 Recall that the quantum
integrability for a lattice model of $N$ sites ensures that, there must exist
independent and mutually commuting set of conserved operators $\{C_j\}, j=1,
\ldots, N,$ the Hamiltonian being just one of them. For achieving this one
usually shows that a one-parameter family of transfer-matrix
$\tau(\la)=tr (\prod_j L_j(\la)$, constructed from the Lax operator $
L_j(\la)$ and generating the set of conserved operators, themselves  commute.
This in turn follows from  the  Yang-Baxter equation
(YBE) satisfied by the associated
$ L_j(\la)$-operator and the $R$-matrix solutions.
Note that, while in  CBA solvable models one can solve the eigenvalue
problem (EVP) only for the Hamiltonian, the quantum integrable systems allow
exact solution  of the  EVP,  simultaneously for the whole set of conserved operators,
including the Hamiltonian. This is done by solving the EVP of the
transfer matrix: $\tau(\la)|n>=\Lambda_n(\la)|n>$ through ABA. Consequently,
for describing quantum integrable systems \c{qis} one has to start not from the
Hamiltonian of the model but from the related  solutions of the YBE.

 Our aim here  is to look
into the class of  solvable 
 electron and spin  models mentioned above, from the view point of quantum
 integrability.  We  therefore start  by discovering the associated
 $R$-matrices and the quantum Lax operators for the integrable
 class in such systems. In doing so we exploit certain symmetries of the YBE
under twisting and gauge transformations. This  establishes in one hand the
complete quantum integrability of the system   
and allows to  apply the ABA method for   exactly solving
 the EVP for the
transfer matrix and consequently   for  all the 
conserved operators, simultaneously.
 At the same time our formulation helps to  identify the integrable
subclass of CBA solvable models  classified in 
\c{oster00}  and 
  find  the precise relationship between various
 existing models \c{borov912,kun98,SS98,oster00}
 at their Lax operator as well as Hamiltonian level.
As an important consequence, we detect some error in the derivation of the
Zvyagin et al  model
 \c{borov912,zvyag92}, derive the correct one and thus completely resolve
the controversy regarding the solvability of the Zvyagin et al  model
and its
 equivalence with the  Schulz-Shastry  model
\c{zvyag99,SS99}. In addition  our approach based on 
the construction of  Lax-operators allows us to
generate new quantum integrable models including generalized $t-J$ models

The organization of our paper is as follows. Sect. 2. describes briefly the
 main
ideas   of quantum integrability and ABA.
Sect. 3 introduces the  symmetries of  YBE and the construction of  the  
 $R$-matrix and the Lax-operator.
Sect. 4 is devoted to the derivation of models  showing their
mutual relations and establishes the equivalence between 
different models. The extension for our formulation to 
 the multi-chain models and their equivalence are  also considered here.
 Sect. 5 generates a new class of 
models including one for the generalised
$t-J$ models.  Sect. 6 presents the exact ABA solutions for the
 models constructed here.
 Sect. 7 is the concluding section. 

\section{ Quantum integrability, conserved operators and ABA}
\setcounter{equation}{0}
  By  quantum integrability we will mean the integrability in the
Liouville sense. For classical models such integrability means the existence of
action-angle variables, when conserved quantities including the Hamiltonian
can be expressed through action variables only. For quantum models,
Liouville integrability similarly demands existence of sufficient number of
mutually commuting conserved operators $\{C_j\}$, the Hamiltonian being one
of them. The aim of the quantum inverse scattering is to solve the
eigenvalue problem  for all these conserved quantities, simultaneously.
Recall in this context that the CBA is designed to solve this problem only
 for the
Hamiltonian of the model.  Therefore quantum integrable systems are
much richer, though more
  involved and demanding than the CBA solvable models
  and as a rule the quantum integrability  does not  follow from  its
solvability by CBA. For describing integrable systems, naturally
 one can not  start   from the Hamiltonian  as
 in CBA, but has to introduce some abstract objects
like the Lax operator $ L_j(\la)$ and the $R$-matrix, which depend on
an extra parameter $\la $ called the spectral parameter \c{qis}. Though the Lax operator
is also a matrix similar to the $R$-matrix, it is a $p\times p$-matrix,
  while the matrix  dimension  of $R$ is  $p^2 \times p^2$. Another
important difference is that,
  the  elements of matrix $ L_j(\la)$ are quantum operators defined at site
$j$ and acting in the corresponding
 Hilbert space, while  the elements of the
$R$-matrix are the usual $c$-numbers.
The matrices are considered here in the fundamental representation of the
underlying algebra with its rank generally coinciding with $p-1$. For
example for $su(2)$, as in case of the spin-$\ha$ chains,
 the Lax operator would be a $2 \times 2$ and the
$R$-matrix a $4 \times 4$ matrix.

  A sufficient condition for the quantum
integrability of the system may be given by 
 the celebrated 
YBE
\be
R_{ab}(\la,\mu)L_{aj}(\la)L_{bj}(\mu)=L_{bj}(\mu)
L_{aj}(\la)R_{ab}(\la,\mu), \ \ \   j=1,2,\ldots,N,
\ll{ybe} \ee
where $a,b$ indicate the matrix or the auxiliary  spaces,
 while $j$ the quantum spaces. The associativity property of the 
algebraic relation (\re{ybe}) yields in turn a similar relation
\be
R_{ab}(\la,\mu)R_{ac}(\la,\gamma)R_{bc}(\mu, \gamma)=R_{bc}(\mu,\gamma)
R_{ac}(\la,\gamma)R_{ab}(\la,\mu),  
\ll{rybe} \ee
representing a compatibility condition for the $R$-matrix.
  The 
$R(\la,\mu)$-matrix usually   depends on the difference between 
 spectral parameters $\la$ and $\mu$, though there may be exceptions as in
the case of Hubbard model. The well known and well studied integrable models
belong to the so called ultralocal class, which along with the YBE
(\re{ybe}) obeys also an additional restriction on the $L$-operators:
 $ [L_{ai} (\lambda),  L_{bj} (\mu)]=0, i \not =j , a \not = b.$ 
Since the notion of  integrability  is  intrinsically related to 
   the conserved quantities, which are global objects concerning the whole
system, for its description we must switch over from the local YBE
(\re{ybe})
 to some global relations. For this purpose therefore, we define a global object known as the
monodromy matrix, by matrix-multiplying the Lax operators:
 $T_a(\la)=\prod_{j=1}^N L_{aj}(\la),$ 
  which yields again a $p \times p$ matrix with operator elements $T_{\alpha
\bet}.$
  Multiplying the local relations
 (\re{ybe}) successively for all sites $j$  we   arrive at 
  the global YBE exhibiting the  same form:    
\be~~~ R_{ab}(\lambda - \mu)~ { T_a} (\lambda)~ { T_b}(\mu )
~ = ~  { T_b}(\mu )~ { T_a }(\lambda)~~ R_{ab}(\lambda - \mu).\ll{ybeg}\ee
Note that the construction of this global YBE
 is possible only due to the ultralocality property,  when
 the Lax operators at different sites can be treated  almost like commuting 
 classical objects and   can be dragged through one another
 and multiplied to yield (\re{ybeg}).  
This algebraic invariance  of the tensor product 
 reflects the deep  Hopf algebra structure underlying  all integrable
systems. Taking now the trace from both sides of relation (\ref{ybeg}),
 the $R$-matrices get canceled,
 since they can be rotated cyclically under the trace
 and  one gets finally for the transfer matrix 
 $\tau(\lambda) =tr (T(\la))$ the crucial relation 
$ [\tau(\lambda),\tau(\mu)]=0 .$  Defining the conserved quantities 
as  the expansion coefficients of 
\be
ln \tau(\la)=\sum_n C_n\la^n,
\ll{cn}\ee
 the commutativity   of the  set  
$\{C_n\}$ for different values of  $n$: $[C_n,C_m]=0, $
  follows immediately and that 
 establishes the 
  complete quantum  integrability of the  system. 
\subsection{Hamiltonian construction}
Though the starting  point in a Quantum integrable system is an abstract
 Lax operator, it can  systematically generate the explicit forms of all
 conserved operators of physical interest including the Hamiltonian of the
system. This follows from the fact that
 the Lax operators through successive multiplication
leads to the monodromy matrix and the trace of it gives the transfer matrix,
which in turn using
  (\re{cn}) generates  the conserved quantities
$C_n= {1 \ov n!} {\partial
\over  \partial \la} ln \tau(\la)\mid_{\la=0}
 $. For example, if the 
 Hamiltonian of the model is defined as 
$ H \equiv C_1= 
 \tau^{'} (0) \tau^{-1}(0)$ and 
the Lax operator satisfies 
an important criterion called regularity condition:
  $  { L}_{aj}(0)={ P}_{aj}$,  linking it with  
  the permutation operator ${ P}_{aj}$, then 
 the  Hamiltonian  may be expressed directly through the Lax operator as
\be 
 H=\sum_j H_{jj+1}, \ \ \ H_{jj+1}= 
{ L}^{'}_{jj+1}(0){ L}^{-1}_{jj+1}(0)= { L}^{'}_{jj+1}(0){ P}_{jj+1}.
\ll{H}.\ee
Note that due to the regularity condition of the Lax operator 
one can  use the  properties of the permutation operator: 
  $  { P}_{aj} { L}_{aj+k}={ L}_{jj+k}{ P}_{aj}$ and $
{ P}_{aj}^2 = tr_a{ P}_{aj}=I $ to
  derive   Hamiltonian (\re{H}), exhibiting    
 nearest neighbour interactions. Fortunately,
fundamental integrable models in condensed matter physics, e.g. spin models,
Hubbard model, $t-J$ model etc. fall into the category of ultralocal
as well as regular model and their Hamiltonians  are with nearest neighbour 
interactions described by $C_1$. The theory of nonultralocal integrable
 models is still in the developmental stage \c{kunhlav} and will not be
considered here.

\subsection {Algebraic Bethe ansatz}
As we have stated above, while   
 the CBA is
concerned  with the exact solution of the  EVP for the Hamiltonian $H,$
the ABA aims to do so   for all
 conserved operators simultaneously. This is achieved 
 by solving the EVP  for   
 the transfer matrix, expressed through the diagonal elements 
   of the monodromy matrix as
 $\tau(\la)=\sum_\alpha T_{\al \al}(\la)$. The off-diagonal elements   
of the monodromy matrix $T_{\al \bet}(\la_i)$ with $\al<\bet,$
 on the other hand correspond 
usually  to the pseudoparticle creation operators and their successive
actions 
on the  pseudovacuum $|0>$ generate the $n$-particle eigenstates 
 $|n>$.
The states $|n>$  and consequently the corresponding  eigenvalues of
$\tau(\la)$   
 depends on $p-1$ different sets of spectral parameters 
$\{\la^{(r-1)}_{i_r} \}, r=1, \dots,p-1$ also known as the rapidity variables,
such that the total number of excitations matches with the pseudoparticle
number:$\sum {i_r}=n. $
 Therefore for calculating  
the related eigenvalue problem, one has to examine the actions of the
 diagonal elements $ T_{\al \al}(\la)|n>$, for which one needs the explicit 
 commutation
relations  between the diagonal and the  off-diagonal elements 
 of $T(\la)$. This knowledge is required for dragging    the former type of
 operators through the later type until they
hit the pseudovacuum state $|0>$. The resulting problem
  $T_{\al \al}(\la)|0>=a_{\al}(\la)|0>$ can be readily solved, since 
$a_{\al}(\la) $ is  obtained from the trivial action of
the Lax operators on the pseudovacuum state.
 Demanding $|n>$ to be the
true eigenstate of $\tau(\la)$, one gets on the other hand
  some important extra conditions,
known as the Bethe equations, for determining the  rapidity variables
$\{\la^{(r-1)}_{i_r} \}$.
The exact procedure of ABA depends naturally on the concrete models.
Recalling  $p-1$ as the rank of the underlying  algebra, we see that 
for systems with $p =2$, e.g. for the spin-$\ha$ chain  there is only
a single set of rapidities $\{\la_{i} \}$ , while  for $m$-chain spin
models, one gets
 $m$ sets of rapidities $ \{  \la^{(r-1)}_{i_r} \}, r =1,\ldots, m.$
Similarly,  
the  Hubbard and the supersymmetric  $t-J$ models are described by 
  two sets of
rapidities $\{\la^{(0)}_{j} ,  \la^{(1)}_{\al} \}.$ 
Note however that the underlying symmetry algebra of the Hubbard model is 
$su(2) \times su(2)$, resulting a   $4\times 4$ Lax operator,
 while that for the SUSY $t-J$ model is $gl(1,2)$
 with its Lax  operator being a $3 \times 3$ matrix.
For models with $p >2$  the ABA steps are recursively repeated
 giving the nested  Bethe
ansatz. We shall give some  details on the twisted Hubbard and the $t-J$
 models
in sect. 6 for elaborating this process.

\section{ Symmetries of  YBE and
Generating transformations}
\setcounter{equation}{0}
For identifying    a  procedure for generating integrable multi-chain electron
 models with correlated hopping
  and  spin models with interchain interactions 
we  intend   to find certain symmetries of the YBE, which would yield  
  under  particular transformation  Lax operators  and $R$-matrices
   as new
 solutions of the
YBE. Therefore the idea is to start with the $L,R$ matrices of known
integrable  systems and derive the transformed ones
  satisfying again the YBE and therefore  representing new integrable
systems. From these transformed Lax operators one can derive now the
Hamiltonians of the desired integrable models with
 correlated hopping.

We look for   the set of transformations 
 given by
\be
 R_{ab}( \la, \mu) \to
 \tilde R_{ab}(\la, \mu)=F_{ab} R_{ab}( \la, \mu)
 F^{-1}_{ba}; \ \   
\tilde L_{aj}(\la)=F_{aj} L_{aj}(\la) F^{-1}_{ja},
\ll{GT}\ee
and demand that the YBE  (\re{ybe})  should remain valid  under such
a transformation. In what follows we will frequently denote $L_{aj} $
by  $R_{aj} $ for convenience. 
It can be checked easily through simple algebra that, 
(\re{GT}) 
maps a solution of the YBE
   into another one, only if the following set of relations holds.
\be
R_{ab}F_{aj}F_{bj}=F_{bj}
F_{aj}R_{ab},\ \
R_{aj}F_{bj}F_{ba}=F_{ba}
F_{bj}R_{aj}, \ \
R_{bj}F_{ba}F_{ja}=F_{ba}
F_{ja}R_{bj},
\ll{ybet} \ee
 along with the  conditions  
that $F_{ab} $  must itself  be a solution of 
  the YBE  (\re{rybe}):
\be 
F_{ab}F_{ac}F_{bc}=F_{bc}
F_{ac}F_{ab}. \ll{fybe}\ee

In our constructions we will explicitly find the transforming operator 
$F$ by assuming certain symmetries.
 For example,  supposing the symmetric
condition
$F_{ba}=F_{ab}\equiv S_{ab}$, transformation (\re{GT})
looks like a gauge transformation
\be
 \tilde R_{ab}(\la, \mu)=S_{ab} R_{ab}( \la, \mu)
 S_{ab}^{-1}; \ \   
\tilde L_{aj}(\la)=S_{aj} L_{aj}(\la) S^{-1}_{aj},
\ll{sGT}\ee
while with anti-symmetry  
$F_{ba}^{-1}=F_{ab}\equiv T_{ab}$, which is  
 known as the twisting transformations
\c{twist} we get 
\be
 \tilde R_{ab}(\la, \mu)=T_{ab} R_{ab}( \la, \mu)
 T_{ab}; \ \   
\tilde L_{aj}(\la)=T_{aj} L_{aj}(\la) T_{aj}.
\ll{tGT}\ee

For  model construction we  consider   transformations with
such symmetries only and satisfy trivially the essential condition
  (\re{fybe})     
 by choosing the transforming operators $F$ as  mutually commuting.
Notice  that 
under  these  conditions   all relations in (\re{ybet}) can be reduced 
for both the  gauge and  the twisting 
transformations
to a single convenient  form
\be  R_{ab}F_{ac}F_{bc}=F_{bc}
F_{ac}R_{ab}, \ll{ybesa}\ee
 where the indices $a,b,j$ are treated in equal footing with $R_{aj}$ to be
understood as  
 $L_aj$.

There exists another  type of  gauge transformation, 
where the  symmetric  operators are
 given   in the factorised form:
 $S_{ab}\equiv G_{ab}= g_ag_b$ . It can be shown easily that, in such cases  
  no extra
condition is imposed for the validity of the transformed solutions and
 the operators  $g_a$  can be arbitrary invertible matrix. To demonstrate this  we may 
focus on the lhs of the YBE (\re{ybe}) and insert the transformed 
solutions
(\re{GT}) with  the   transforming  operators in the factorised  form
to get
\bea
\mbox {lhs} 
&=&g_ag_bR_{ab}(\la,\mu)g^{-1}_ag^{-1}_bg_ag_jL_{aj}(\la)g_a^{-1}g_j^{-1}
g_bg_jL_{bj}(\mu)
g_b^{-1}g_j^{-1}\nonumber \\&=&
g_ag_bg_j\left (R_{ab}(\la,\mu)L_{aj}(\la)L_{bj}(\mu) \right)
g_a^{-1}g_b^{-1}g_j^{-1}.
\ll{lybe} \eea
Similar calculations are repeated   for the rhs and since 
the old $ R, L$ solutions satisfy the YBE,
 it  holds naturally  for the transformed solutions. Note that in dragging the
factors involving $g_a$'s out of the original lhs of YBE as done in
(\re{lybe}), we have used only the mutual cancellation of these factors as
well as trivial commutativity of the operators acting on different spaces.
For such transformations as evident from (\re{lybe}) the validity of the
YBE condition
(\re{fybe}) is also not required.
 Therefore the choice of the
transforming operators $g_a$ can be completely arbitrary. We show below that
using such freedom we can construct integrable models which can go beyond the
class of models classified in \c{oster00}.  However confining only to
mutually commuting   
twisting and gauge transformations, as mentioned 
 above, we can derive an integrable  subclass 
of the solvable models considered in  \c{oster00}. We emphasise again
that
 all the models
we generate here  belong to the quantum integrable systems and therefore
are much richer than the models, which allow only the 
solution of the  Hamiltonian eigenvalue  problem through  CBA.

\subsection{Transformed Hamiltonian}
Recall that 
  (\re{H}) links Hamiltonian of a model directly to its Lax operator,
if  the regularity condition is  fulfilled. 
We notice further that
 the regularity condition is preserved under  transformations (\re{GT}),
since it gives
$\ti L_{aj}(0)= F_{aj}L_{aj}(0)F_{ja}^{-1}= F_{aj} P_{aj}F^{-1}_{ja}
= F_{aj} F^{-1}_{aj}P_{aj}=P_{aj}.
 $
Therefore one gets  a precise way for obtaining
   new transformed Hamiltonians  explicitly
from the transformed Lax operators (\re{GT}) as
\bea H&=&  \sum_j
{\tilde L}^{'}_{jj+1}(0){\ti L}^{-1}_{jj+1}(0),\nonumber \\
&=& \sum_j ( 
F_{jj+1}L_{jj+1}^{'}(0)F_{j+1j}^{-1})(F_{j+1j}
L^{-1}_{jj+1}(0)F^{-1}_{jj+1}),\nonumber \\
&=& \sum_j 
F_{jj+1}(H_{jj+1})F^{-1}_{jj+1},
\ll{HZ} \eea 
where $H_{jj+1}$ corresponds to the original model as defined in (\re{H}).
Therefore if one starts from  known integrable models like $XXZ$ or Hubbard
model, with their Lax operators satisfying both ultralocality and regularity
conditions, then through (\re{GT}) one can derive the transformed Lax
operator with the same essential properties and in turn can generate through
(\re{HZ}) the  Hamiltonian with new interactions   introduced by the operator
$F_{jj+1}$.

\section{Model construction}
 \setcounter{equation}{0}
For constructing  new integrable models following the above scheme
we  restrict ourselves only to the  symmetric (\re{sGT}) 
 and antisymmetric (\re{tGT}) transformations and 
  build the transforming   operators  out of the Cartan generators
$H_{(\al )},
\al =1, \ldots,r,$ $r=p-1$ being the rank of the underlying  algebra.
Due to the mutual commutativity of $H_j$  the  transforming operators
 would automatically satisfy the YBE (\re{fybe}) and
therefore  the  only  condition that 
 remains to be satisfied for our 
transformations $S_{ab}$ and $T_{ab}$ is  (\re{ybesa}).  
It can be shown  that such operators in the general case may be
constructed in the exponential form \c{twist}
\be
{\sc G}_{aj}= e^{i
\sum_{\al \bet} \theta_{\al \bet} (H_{a(\al )} H_{j(\bet)})},
\ll{G} \ee
with symmetric or antisymmetric properties  on the parameters: $\theta_{ \bet \al}=\pm 
\theta_{\al \bet}$.
In (\re{G}) $\al =0$ may also be included in the sum by
 considering $H_0 \equiv 1$.
Note that the rank $r$ of the associated algebra depends on the concrete
models and determines also the  matrix  dimension of the Lax operator 
together with  the chosen  representation. However since  we are interested
here in a particular class of models  with relevance in 
 condensed matter physics, e.g.  spin chains, Hubbard model, t-J
model etc. we would prove the validity of (\re{G}) in such particular
cases only. 

 \subsection{Correlated electron and spin ladder models}

We consider first the   Hubbard like 
correlated electron models  involving operators 
$  c^\dag_{j(\al)},c_{j (\al)},
  n_{j  (\al)}$ with spin components 
$\al =\pm$ and $XXZ$ or $XXX$ type two-chain   or the so called ladder models 
 involving two independent spin-$\ha$ operators $ \vec\sigma, \vec \tau$.
Subsequently, we will look into   
 their multi-chain generalisations as well as transformed   $t-J$ and some other new
 spin  models. Note  that
since through Jordan-Wigner transformation the spin operators can be mapped
into fermions and vice versa, the Lax operators 
 in case of  spin  models can be obtained analogous to those for
the fermionic models, which we consider here in details.
 Therefore in conformity with (\re{G}) 
we propose the symmetric gauge transformations
  as
\be
S_{aj}(s)= e^{i s (n_{a(-)} n_{j(+)} + n_{a(+)} n_{j(-)} )}.\ll{SG} \ee
while the antisymmetric  twisting transformation in 
the   form
\be
T_{aj}(\theta, \gamma_{(\pm)})= e^{i[ \theta (n_{a(-)} n_{j(+)}
 - n_{a(+)} n_{j(-)} )+\gamma_{ (+)}( n_{a(+)}- n_{j(+)})+
\gamma_{ (-)}(n_{a(-)}- n_{j(-)})},
\ll{T} \ee
As we have stated above, 
  we have to check only  the validity of 
(\re{ybesa}) for both these operators to  confirm their usefulness for
generating new integrable systems. In all our calculations we need only
 to use
 the following basic   commutation relations  for  fermions
 or  spin
operators like 
\be f(n_{a(\bet)}) c^\dag_{j(\al)}=\delta_{aj} \delta_{\al \bet}
c^\dag_{j(\al)}f(n_{a(\bet)}+1), \ \ 
f(n_{a(\bet)}) c_{j(\al)}=\delta_{aj} \delta_{\al \bet}
c_{j(\al)}f(n_{a(\bet)}-1)\ll{crf}\ee
\be
[\sigma^\pm,\sigma ^3]=\mp 2\sigma ^\pm  , \ \  
[\sigma^+,\sigma^-]=\sigma^3,\ \ (\sigma^\pm)^2=0,
\ll{cr} \ee
etc.
We note first that the $R_{ab}$-matrix or the Lax operator 
of the Hubbard model (similarly those for the $XXZ$ or $XXX$ spin models)
contains strings of   operators in the quadratic form $ c^\dag_{a(\pm)}c_{b (\pm)}$,
which   are the only parts  not commuting with the operators $n_{a(\pm)} $
forming the above transformations.
 However it is 
 evident  that, any function of  the total number
operator $n_{a(\pm)}+n_{b(\pm)}$  commutes with this quadratic form and
hence   with the  $R_{ab}$, while  the  operator
 $n_{c(\pm)} $  commutes
trivially with it. We note now that the transforming matrices in (\re{ybesa}) induce
operators exactly in such
  combinations:
$ F_{ac}F_{bc} \sim e^{i const. \left((n_{a(-)}+n_{b(-)}) n_{c(+)} \pm 
( n_{a(+)} +n_{b(+)}) n_{c(-)} \right)}$,
 for (\re{SG}) and  (\re{T}),  respectively, which 
 based on our above arguments  proves  the validity 
of 
(\re{ybesa}) and therefore the validity of (\re{GT})  for both (\re{SG})
 and (\re{SG}). 
 We may consider also an additional transformation 
in the factorised form 
choosing 
\be
G_{aj}(g)=g_ag_j \ , \ \  g_{a}(g)= e^{i  g (n_{a(-)} n_{a(+)} )}.
\ll{Gt} \ee
Since  (\re{SG}), (\re{T}) and (\re{Gt})
can be applied independently, we may consider their combined effect 
by their successive application: 
\be F_{aj}(\theta,\gamma_{(\pm)},s,g)=S_{aj}(s)T_{aj}(\theta,\gamma_{(\pm)})G_{aj}(g),
\ \ \ti F_{aj}(\theta,\gamma_{(\pm)},s,g)=S_{aj}^{-1}(s)T_{aj}(\theta,\gamma_{(\pm)})(G_{aj})^{-1}(g)
\ll{asg} \ee
for constructing transformation through 
 (\re{GT})  as
\be
\ti R_{aj}( \th ,\gamma_{(\pm)} , s,g, \la) = F_{aj}( \th ,\gamma_{(\pm)} , s,g)R^{(Hub)}_{aj}(\la)
 \ti F_{aj}(\th ,\gamma_{(\pm)} ,s,g). \ll{Lt}\ee
Here the transformed $\ti R$ in (\re{Lt})  stands for both
the changed Lax operator and the  $R$-matrix and    
 naturally satisfies the YBE  representing 
a new quantum integrable system with $N$ number of independent conserved
operators. These systems would  give
a hierarchy of extended Hubbard models with correlated hopping or 
through  Jordan-Wigner mapping from fermion to spin operator:
 $n_{a(\pm)} \rw 
{1\over 2}
(1+\sigma^3_{a(\pm)})$ and $ c_{j (\pm)}\rw \sigma^+_{j(\pm)}, \ 
c^\dagger_{j (\pm)}\rw \sigma^-_{j(\pm)}, $
a hierarchy of   spin ladder models with interchain interactions.
We have used here the notation $\vec \sigma_{(-)}$ in place of $\vec \tau$ for
convenience.

 The Hamiltonian of such transformed Hubbard systems would be given by  
 (\re{HZ}) with the use of the transformation (\re{asg}) as  
\bea \ti H&=&\sum_j F_{jj+1}( \th ,\gamma_{(\pm)},s,g)
H^{(Hub)}_{jj+1}F^{-1}_{jj+1}( \th ,\gamma_{(\pm)},s,g)
\nonumber \\
&=& -\sum_{j}  c^\dag_{j+1 (+)}c_{j (+)} h_{jj+1}^{(+)}
+  c^\dag_{j+1 (-)}c_{j (-)} h_{jj+1}^{(-)}
+ U  n_{j  (+)}n_{j (-)}  +{\rm h.c}
, \ll{Hubt} \eea
exhibiting explicit correlated hopping 
\be
 h_{jj+1}^{(\pm)}=
e^{i \left(-2\gamma(\pm) 
+(\pm \th - s+g) n_{j+1(\mp)}+(\pm \th + s-g) n_{j(\mp)}\right)}.
\ll{S}\ee
Its different  parameters  keep track of the independent
 transformations (\re{SG}), (\re{T}), (\re{Gt}) and therefore one can study
each or any combinations of them by switching off the  others. In this way,
as we see below, one can identify different known models showing also the
equivalence between them.  

\subsection{Connection with general setting and equivalence}

We can now link our model  to the   correlated hopping
 electron models classified in 
\c{oster00}.
This 
comparison reveals that our model described  by the 
  Hamiltonian  
 (\re{Hubt}) with (\re{S})  constitutes 
 an important  integrable subclass of  the  solvable
  models of \c{oster00}, defined  by a particular choice
of their coefficients $\al_{jm}(\sigma) $ and $\gamma_j(\sigma) $ 
appearing  in the 
correlated hopping terms 
 as 
\be
\al_{jm}(\sigma)=\delta_{jm} (\sigma\th +s-g)+
\delta_{j+1m}(\sigma \th   -(s-g)), \ \ \ 
 \gamma_j(\sigma)=-2 \gamma_{(\sigma)}. 
\ll{link}\ee
where $\sigma =\pm .$
Therefore the global unitary transformation defined in  
\c{oster00}  \be 
U=exp \left(i\sum 
(\xi^{\al, \bet}_{j,k} n_{j(\al)}n_{k(\bet)} +\zeta_{j,\al}
n_{j(\al)})\right) .
\ll{U}\ee
that introduces correlated hopping  in  the Hamiltonian
can also be identified for our case   by equating the coefficients as 
\be
\xi^{\sigma,- \sigma}_{j,j}=
\xi^{\sigma,- \sigma}_{j+1,j+1}=-\ha (s-g), \ \ 
\xi^{\sigma,- \sigma}_{j,j+1}=-
\xi^{\sigma,- \sigma}_{j+1,j}= \ha \sigma \th 
\ll{xi}\ee
 with the rest of the coefficients
 in (\re{U}) including all $\xi^{\sigma,\sigma}_{j,k} $  being zero.
As a consequence  all coefficients $ A_{j,k}(\sigma)=2(
\xi^{\sigma,\sigma}_{j,k} -\xi^{\sigma,\sigma}_{j+1,k}),\ 
 k\neq (j,j+1) \  $ 
appearing  in the models of 
 \c{oster00} also vanish. Moreover it is significant  to note  
 that the class of quantum integrable 
models we construct correspond to the 
 translational invariant subclass of the solvable models \c{oster00}
, as is evident from (\re{link}),
 since we have here
$\al_{jm}(\sigma)=\al_{j-m}(\sigma)  $ with its only nontrivial coefficients
$\al_{0}(\pm)=\pm \th +s-g $ and $
\al_{-1}(\mp)=\pm \th   -(s-g), $
satisfying the necessary condition $\al_{0}(\pm)+\al_{-1}(\mp)=0 $.
Weather this linkage between the quantum integrability and the 
translational invariance has any deeper meaning is yet to be explored.
On the other hand
we recall  that the  models introduced by Schulz-Shastry 
 \c{SS98}
also exhibits  translational  invariance, since their
   coefficients correspond to  $\al_{jm}(\pm)=\mp
\al_{j-m} $ with $A_{j,k}(\sigma)=\gamma_{j(\pm)}=0$. 
Comparing with (\re{link}) we may  notice 
that, the Schulz-Shastry model, in spite of its translational invariance,
 does not  in general correspond to our quantum integrable
case. However,  we will see below that   a further restricted
 model, known as the {\it minimal} Schulz-Shastry model,
 given by the particular choice
$ \al_{0}=\al_{-1} \neq 0$ with all other $\al_{k}=0$ does agree
 with a particular
case of the present construction, showing that the 
{\it minimal} Schulz-Shastry model belongs to a quantum
integrable class.  
 Returning again to the present case, we 
 observe that just by  restricting
 to the choice (\re{link}) and (\re{xi}),
  we can   use all the formulas derived in  \c{oster00} for our
model  and thus 
 can calculate  the set
of boundary phases and other relevant objects for the present case.
 Moreover, we can find that the necessary condition for the CBA solvability
 of such models prescribed in \c{oster00},
  holds also for our model, since  for the choice
(\re{link}) it is  easy to check that the compatibility condition:
$\al_{jm}(\sigma)+\al_{mj+1}(-\sigma)
-\al_{jm+1}(\sigma)-\al_{mj}(-\sigma)=0 $ with $m=j,j+1$ only, holds true.

Therefore we conclude that,
 while all Amico et al model \c{oster00} are only CBA solvable, their
particular subclass
defined by 
(\re{link}) and (\re{xi})
and
 generated by our construction 
  not only becomes CBA solvable but also represents an
 exact  quantum integrable
system. This integrable subclass may be  given by the
Hamiltonian (\re{Hubt}) with (\re{S}) together with all its particular cases
and is associated 
 with explicit Lax operator and $R$-matrix solutions (\re{Lt}).
Consequently, these models
 allow a hierarchy of mutually commuting conserved operators
having
 increasingly  further neighbour interactions, e.g.
 nearest, next nearest, second next
 nearest neighbours  etc., as well as more and more nonlinear
 interacting terms in their successive  Hamiltonians.
Moreover, as we have mentioned already, while CBA aims to solve the
eigenvalue problem for the Hamiltonian of a single model, quantum
 integrable systems
permit exact
   eigenvalue  solution  for all its    $N$-number  of      
 conserved   operators,         simultaneously  through
ABA. We will demonstrate briefly this ABA application to 
our  integrable model   (\re{Hubt}),(\re{S}) in sect. 6.

We stress again that one can generate various extensions to the Hubbard model
from  (\re{Hubt}) by  
 adjusting the independent parameters
$\gamma_{(\pm)}, \th ,s,g$ in (\re{S}), all of
  which would be  
   quantum integrable systems allowing application of ABA, as well as
solvable through CBA models. Their Lax
 operators and $ R$-matrices   would naturally be related to each other by
gauge or twisting transformations
 giving  a direct  way of establishing  explicit relationship between the
models
 with different types of  hopping interactions    at      their      Lax
 operator level. This throws a new light on the question of equivalence of
 the models, which created significant controversy in recent years
\c{SS99,zvyag99,oster00}.   
For example, 
if we restrict           only to the
twisting        transformation by choosing    $\th \not =0,$ while switching
off  the other  parameters:
 $   g=s=\gamma_{(\pm)}=0,$ 
          we get an   integrable model    with  the Lax    operator
\be
\ti R_{aj}( \th , \la) = T_{aj}( \th,0 )R^{(Hub)}_{aj}(\la)
T_{aj}(\th,0) \ll{LLt}\ee
   with    
  $T_{aj}(\theta, 0)= e^{i[ \theta (n_{a(-)} n_{j(+)}
 - n_{a(+)} n_{j(-)} )}
$ as    in  (\re{T}) and 
    the      corresponding      Hamiltonian    
will be  given    as
(\re{Hubt})      with its hopping interaction reduced accordingly  from
 (\re{S}) as
\bea \ti H 
&=& -\sum_{j} ( c^\dag_{j+1 (+)}c_{j (+)} e^{ i \th 
(  n_{j(-)}+ n_{j+1(-)})}
+  c^\dag_{j+1 (-)}c_{j (-)} e^{-i \th (  n_{j(+)}+ n_{j+1(+)})}
\nonumber \\ 
&+& U  n_{j  (+)}n_{j (-)}  +{\rm h.c.})
 \ll{Hk} \eea
We     observe      that     (\re{LLt})    coincides with the $L$-operator
found in \c{kun98}, while  (\re{Hk}) 
recovers the Hamiltonian of the Kundu  model  \c{kun98}, which also 
 tallies  interestingly
 with the {\it minimal} Schulz-Shastry model \c{SS98,SS99}. Clearly,
   from (\re{Hk}) one  derives
$\al_{0}=\al_{-1}=\th$ as the only nonvanishing coefficients in its
correlated hopping term, which   defines 
 the {\it minimal} Schulz-Shastry model,  as mentioned above. 
  We therefore conclude that the {\it minimal} Schulz-Shastry  model
$H^{(min-SS)} $, given by the
Hamiltonian (\re{Hk}) and
  identical to the  model of  \c{kun98} 
 represents a quantum integrable system with its  Lax operator and the
 $R$-matrix $R^{(min-SS)}_{aj}( \th ,
  \la)$ given by (\re{LLt}).

Let us  consider now a combination of the gauge and the twisting
transformation by adjusting the corresponding parameters as
 $s=\th, $  while switching off the others by 
  choosing  $ g=\gamma_{(\pm)}=0. $
 The resulting  integrable
system  will have   the   Lax   operator and the $R$-matrix  
 (\re{Lt}) with the transforming operators
$ F_{aj}(\theta)=S_{aj}(\th )T_{aj}(\theta,0)=
e^{2i \theta (n_{a(-)} n_{j(+)}))},
\ \ \ti F_{aj}(\theta)=S_{aj}^{-1}(\th)T_{aj}(\theta,0)=
e^{-2i\th (  n_{a(+)} n_{j(-)})},
$  expressed in the form   
\be
\ti R_{aj}( \th , \la) = 
e^{2i \theta (n_{a(-)} n_{j(+)}))}R^{(Hub)}_{aj}(\la)
e^{-2i\th (  n_{a(+)} n_{j(-)})}, \ll{Lts}\ee
    for      the     
extended    Hubbard  model     and       similarly 
\be
\ti R_{aj}( \th , \la) = 
e^{2i \theta (\tau^3_{a}
\sigma^3_{j})}
(R^{(xxz)\sigma}_{aj}(\la)R^{(xxz)\tau}_{aj}(\la))
e^{-2i\th ( \sigma^3_{a}\tau^3_{j})}, \ll{Lts1}\ee
   for
the XXZ ladder model. It is exciting to  recognise that   Lax operator
(\re{Lts}) (or (\re{Lts1})) is
 exactly same as that for the extended 
 Hubbard model $R_{aj}^{(Z)}( \th,  \la)$ constructed by  Zvyagin et al
  \c{borov912,zvyag92} for their single chain case.
  Comparing  (\re{Lts})     and (\re{LLt})  we observe therefore that  
these Lax
operators are related only by an unitary
transformation and consequently we come to the
       important conclusion that the Zvyagin et al model is a quantum
integrable model and is gauge equivalent to the {\it minimal} Schulz-Shastry 
 model at the Lax
operator level, i.e. we have   \be R_{aj}^{(Z)}( \th,  \la)
=S_{aj}(\th)R^{(min-SS)}_{aj}( \th ,
  \la)S^{-1}_{aj}(\th)
,\ \ \  \  S_{aj}(\theta)= e^{i[ \theta (n_{a(-)} n_{j(+)}
 + n_{a(+)} n_{j(-)} )}.
 \ll{z-ss}\ee
Moreover since the Hamiltonian  can be    determined
uniquely  from the Lax operator 
 as shown in (\re{H}), the equivalence 
 between the above two models extends also  to
 their Hamiltonian level.  However,
  this equivalence  remained obscured \c{SS99} possibly due to the fact 
that in spite of the correct form of the Lax operator in \c{borov912,zvyag92}, the
corresponding 
 Hamiltonian 
  as presented in the same papers          \c{borov912,zvyag92}
 involves some error
\c{private}. As a consequence  this form of the Hamiltonian  
appears to be  even nonsolvable by CBA, as observed in  \c{oster00}, whereas 
being associated with  solutions of YBE it should certainly be 
an exact quantum   integrable model! Our approach resolves this puzzle by
deriving  
the correct form of the  Hamiltonian  from 
(\re{HZ}) using the 
 Lax
operator 
$R_{aj}^{(Z)}( \th,  \la)$ of \c{borov912,zvyag92}, which coincides with 
 (\re{Lts}). Therefore one 
concludes that  the correct Zvyagin et al  model 
$H^{(Z)} $  should be  given by 
   (\re{Hubt})  through
a reduction of the  parameters 
 $g=\gamma_{(\pm)}=0, s=\th$  in the  hopping interaction  (\re{S})
 yielding
\bea \ti H&=&\sum_j e^{2i \theta (n_{j(-)} n_{j+1(+)}))}
H^{(Hub)}_{jj+1}e^{-2i \theta (n_{j(-)} n_{j+1(+)}))}
\nonumber \\
&=& -\sum_{j}  c^\dag_{j+1 (+)}c_{j (+)} e^{2i \th  n_{j(-)}}
+  c^\dag_{j+1 (-)}c_{j (-)} e^{-2i \th  n_{j+1(+)}} 
+ U  n_{j  (+)}n_{j (-)}  +{\rm h.c}.
 \ll{Hzvyg} \eea

It follows therefore 
from the above arguments that, the proper 
 Hamiltonian of the Zvyagin et al model, which should be 
given by           (\re{Hzvyg})
 is related  through unitary transformation 
to that of the {\it minimal} Schulz-Shastry  model, which
 coincides  with (\re{Hk}). That is one should have
$H^{(Z)}=U_1H^{(min-SS)}U_1^{-1}$, where the global unitary 
 operator $U_1$ can 
 be constructed easily from (\re{U}) and (\re{xi}) 
as $U_1=exp \left({-i } \th \sum_{j} 
( n_{j(+)}n_{j(-)})\right) ,$ which establishes the gauge equivalence
between these two models also at the Hamiltonian level. 
An       alternative           
 choice of  $s=-\th,$ leads           
to a    complementary  model   with Hamiltonian 
(\re{Hubt}) and  correlated hopping     
 $
 h_{jj+1}^{(+)}=
e^{2i \th  n_{j+1(-)}}$
and
 $
 h_{jj+1}^{(-)}=
e^{-2i \th  n_{j(+)}}$, which again is  
  gauge equivalent to the {\it minimal} Schulz-Shastry  model and
can be taken also as an alternative but  correct  form of the 
Zvyagin et al model.
 Since both of the above models are obtained as particular cases of 
(\re{Hubt}), they must be 
 CBA solvable 
as well as   quantum integrable systems.
Based on our explicit results derived below, we may draw another important
conclusion   that, contrary to the Hamiltonian of the models
 the Bethe ansatz results 
are insensitive to the gauge transformations S (\re{SG}) and G (\re{Gt})
and depend only on the twisting transformation T (\re{T}).
 Therefore, since  at the Lax
operator level  the
Zvyagin et al model is S-gauge related   to 
 the  {\it minimal} Schulz-Shastry model,  which again is identical to the 
Kundu model,  for all these three models    the scattering
matrix and the Bethe equations related to their eigenvalue problem
 turn out to be
identical. This  is   explicit from the identical  Bethe ansatz equations 
 presented in respective works
\c{borov912,zvyag92,SS98,kun98}
and possibly  this is the reason why the final Bethe ansatz results in 
\c{borov912,zvyag92} 
turn out to be correct in spite of their incorrect form of the  Hamiltonian.    
\subsection {Multi-chain generalization}
In the above models we have taken $\al =\pm$ for denoting  the 
$\uparrow, \downarrow$   components of a $s=\ha$ single chain 
electron or the 
two independent spin-$\ha$ operators in a spin ladder model.
 For generalizing these models to M-chains we have to 
 consider
 the set of 
 fermionic operators
$  \{ c^\dag_{j(\al)},c_{j (\al)} ,
  n_{j  (\al)}\}$
or spin operators $\vec \sigma_{(\alpha)}$ extending the range of $\al$ to 
 $\al = 1, \ldots, M$.
Note that we can incorporate easily  multi-chain electron models with spin-$
\ha$ by redefining the indices  
 through a break up of  
$\alpha=(\gamma,\pm),\gamma= 1, \ldots, {M \ov 2}$,
  resulting ${M \ov 2}$
number of electron chains. Since the rest of the arguments go parallelly,
 we will not distinguish  spin-less fermions from fermions
with spins in what follows,  for
convenience. Moreover since the idea of construction described above for the
single chain models allows simple generalization  to 
 the multi-chain case, we will give only brief account of such constructions
skipping the details and 
stressing only  the differences.
 Firstly for generalising  the above gauge and twisting 
 operators (\re{SG}), (\re{T}) and (\re{Gt})
 to the  $M$-chain models,  we have to use now  more general
transformations (\re{G})
 to get the transforming operators as
\be
S^{(M)}_{aj}(\hat s)= e^{i\sum_{\al < \bet} s_{\al \bet} (n_{a(\al)} 
n_{j(\bet)} + 
n_{a(\bet)} 
n_{j(\al)} )},\ll{SGm} \ee

\be
T^{(M)}_{aj}(\hat \theta,\vec \gamma , \hat \gamma)= 
e^{i(\sum_{\al < \bet} \th _{\al \bet} (n_{a(\al)} 
n_{j(\bet)} -
n_{a(\bet)} 
n_{j(\al)}) +\sum_\al
\gamma_{ (\al )}( n_{a(\al )}- n_{j(\al)}))}
\ll{Tm} \ee
and
\be
G^{(M)}_{aj}(\hat g)=g^{(M)}_ag^{(M)}_j \ , \ \ 
 g^{(M)}_{a}(\hat g)= e^{i 
\sum_{\al < \bet} g_{\al \bet} (n_{a(\al)} 
n_{a(\bet)}  )},
\ll{Gtm} \ee
 respectively, considering the parameters $\hat s, \hat \th, \hat g$ to be
symmetric under interchange of indices, i.e. $\th _{\al \bet}=\th _{ \bet
\al} $ etc. The Lax operator and the $R$-matrix  of the new
quantum integrable multi-chain model can be constructed generalizing
(\re{Lt}) as
\be
\ti
 R_{aj}^{(M)}(  \hat \th , \vec \gamma, \hat s, \hat g, \la)= 
F^{(M)}_{aj}( \hat \th , \vec \gamma, \hat s, \hat g )
 L^{(M)Hub}_{aj}(\la)
\ti F^{(M)}_{aj}( \hat \th , \vec \gamma, \hat s, \hat g ),
  \ll{mLt}\ee
 where 
the transforming operators  (\re{asg}) are generalised by using  the
combinations of
multi-chain  twisting and gauge transformation 
 (\re{SGm})-(\re{Gtm}) as 
\be F^{(M)}_{aj}(\hat \theta,\vec \gamma, \hat s, \hat g )
=T^{(M)}_{aj}(\hat \theta,\vec \gamma)
S^{(M)}_{aj}(\hat s)G^{(M)}_{aj}(\hat g), \ \
\ti F^{(M)}_{aj}(\hat \theta,\vec \gamma, \hat s, \hat g )
=T^{(M)}_{aj}(\hat \theta,\vec \gamma)
(S^{(M)}_{aj}(\hat s)G^{(M)}_{aj}(\hat g))^{-1}.
 \ll{FM} \ee
 The set of conserved operators,
 the Hamiltonian being  first in this set, can be
constructed in  explicit form
 through a
straightforward generalization of (\re{HZ}) for the $M$-chain case
 by using the transforming operators from (\re{FM}). Since
the transformations
 considered here do not affect the interacting  term in the Hamiltonian
 involving Cartan
generators, we present only its transformed
 hopping or the $XY$ terms given as
\bea \ti H^{(M)}&=&\sum_j^N F^{(M)}_{jj+1}(\hat \th ,
\vec \gamma, \hat s ,\hat g)
(H^{(M) XY}_{jj+1})
(F^{(M)}_{jj+1}(\hat \th , \vec \gamma,  \hat s ,\hat g))^{-1}
\nonumber \\
&=& -\sum_{j, \al}^{N,M}  c^\dag_{j+1 (\al)}c_{j (\al)} h_{jj+1}^{(\al)}
 +{\rm h.c}
, \ll{Hubtm} \eea
with  correlated hopping involving all other chains: 
\be
 h_{jj+1}^{(\al )}=
e^{i \left(-2\gamma(\al ) 
+\sum_{\bet \neq \al} ( sgn (\al -\bet)\th _{\al \bet}( n_{j+1(\bet)}
+ n_{j(\bet)}) 
+(s_{\al \bet}-g_{\al \bet})( n_{j(\bet)}
- n_{j+1(\bet)})
\right)}.
\ll{Sm}\ee
Comparing   with  the classification  of \c{oster00} we
find that,  our Hamiltonian (\re{Hubtm}) is obtained 
for the  particular choice of
their parameters 
\be
A_{jm}^{\al,\bet}
=\delta_{jm} ( sgn (\al -\bet)\th _{\al \bet} +s_{\al \bet}-
g_{\al\bet})+
\delta_{j+1m}( sgn (\al -\bet) \th _{\al\bet} 
  -(s_{\al\bet}-g_{\al\bet})), \ \ \  \gamma_j(\alpha)=-2 \gamma (\al ), 
\ll{mlink}\ee
with $\bet \neq \al, $ which may be considered as the 
$M$-chain generalization of the relation (\re{link}).
Consequently the only nontrivial coefficients appearing in the corresponding
 global unitary transformation (\re{U}) that introduces correlated 
hopping  in the Hamiltonian
   would be given by 
\be
\xi^{\al,\bet}_{j,j}=
\xi^{\al,\bet}_{j+1,j+1}=-\ha (s_{\al,\bet }-g_{\al,\bet }), \ \ 
\xi^{\al,\bet}_{j,j+1}=-
\xi^{\al,\bet}_{j+1,j}= \ha sgn (\al -\bet )\th _{\al,\bet },
\ll{mxi}\ee
as generalisation of (\re{xi}).
The compatibility of the  coefficients (\re{mlink}) with
 the necessary condition for the  CBA solvability  \c{oster00} can be
easily checked.  
Therefore  we may conclude that a subclass of the
 CBA solvable multi-chain  models of \c{oster00} defined through 
 the parameter choice
(\re{mlink}) and (\re{mxi}) should 
also  belong to a much richer class of  quantum integrable systems. 
Restricting to  the twisting transformation 
alone, i.e. considering only the nontrivial parameter choice 
$\hat \th \neq 0,$ we get a model with  
  Hamiltonian (\re{Hubtm}) and the correlated  
 hopping term reduced from  (\re{Sm}) to
$
 h_{jj+1}^{(\al )}=
e^{i \left( 
+\sum_{\bet \neq \al} ( sgn (\al-\bet)\th _{\al \bet}( n_{j+1(\bet)}
+ n_{j(\bet)}) 
\right)}. $  This case is evidently   identical to a subclass  of the 
multi-chain  Schulz-Shastry model 
   \c{SS99}, known as the multi-chain {\it nimimal}
  Schulz-Shastry    model, 
 defined for the only nontrivial choice of their generating function:
 $A_{\al\bet}(0)=A_{\al\bet}(-1)=\th_{\al \bet}$.
At the same time, similar to the single chain case,
 this {\it minimal} class tallies also with
 the multi-chain generalization of the Kundu model introduced
in \c{k00}, where however the interchain couplings $\hat \th$ were 
 restricted only  to   nearest neighbouring chains.
  Therefore we see that
among the multi-chain Schulz-Shastry models only its
 {\it minimal} subclass    exhibits quantum
 integrability and allows
 sufficient number of mutually commuting higher conserved
operators together with 
the  Lax operator and the $R$-matrix  given by
 (\re{mLt}) transformed  by  twisting transformation (\re{Tm}) with 
parameters $\vec \gamma=0$.

 On the other hand
considering a combination of twisting (\re{Tm}) and gauge transformation
(\re{SGm}) by choosing $\hat s=\hat \th, \hat g=0$
one gets  the associated Lax operator from  
 (\re{mLt}) as
\bea
\ti L_{aj}^{(M)Z}( \th , \la)& = &S_{aj}^{(M)}(\hat \th )T^{(M)}_{aj}(\hat \theta,0)
 L^{(M) XY}_{aj}(\la) (S_{aj}^{(M)})^{-1}(\hat \th)T^{(M)}_{aj}(\hat \theta,0)
\nonumber \\ &=& 
e^{2i \sum_{\al < \bet} \th _{\al \bet} (n_{a(\al)} 
n_{j(\bet)}} \left ( \prod_\al ^M L^{(\al) XY}_{aj}(\la)\right) 
e^{-2i \sum_{\al < \bet} \th _{\al \bet}n_{a(\bet)} 
n_{j(\al)} )}.
  \ll{mZLt}\eea
We notice immediately that 
the Lax operator (\re{mZLt}) is a generalisation of  (\re{Lts}) and
coincides  exactly  with that of  
 the multi-chain  Zvyagin et al model \c{borov912,zvyag92}.
 Therefore the Hamiltonian generated from 
it should have the form
\be \ti H^{(M)Z}
=-\sum_{j, \al}^{N,M}  c^\dag_{j+1 (\al)}c_{j (\al)} e^{2i ( 
-\sum_{\bet > \al}  \th _{\al \bet} n_{j+1(\bet)}
 +\sum_{\bet < \al}  \th _{\al \bet} n_{j(\bet)})}
 +{\rm h.c}. \ll{ZHubtm} \ee
This also establishes,
as in the single chain case, that the correct form of the multi-chain Zvyagin
et al  model should be given by the Hamiltonian 
(\re{ZHubtm}), which is related to the {\it minimal}
 multi-chain Schulz-Shastry model
by 
 an  unitary transformation 
$U_M=exp \left(-i\sum_{j,\al \neq \bet} 
\th _{\al, \bet} n_{j(\al)}n_{j(\bet)} \right) ,
$ while the Lax operators of these models are connected through the 
gauge transformation (\re{SGm}) with $\hat s=\hat \th $.  
 This  therefore
 resolves completely the controversies
 raised in all earlier works
\c{zvyag99,SS99,oster00}  concerning the multi-chain case. Restricting to
 $M=1$ one naturally recovers  all results related to the single chain
case obtained above.

\section {Application of the scheme for generating
  new integrable models}
\setcounter{equation}{0}

Using the symmetries of the YBE, as we have shown above, one can generate new
solutions of YBE and hence generate new
quantum integrable models starting from the old ones. Through this scheme 
we have already 
 identified the   integrable 
  subclass of the CBA solvable models  classified earlier.
  Using other options of the same scheme we intend  to generate now new 
models. Firstly we  exploit the freedom in 
 choosing the transformation operator $g_a$ (\re{Gt}) in any arbitrary form.
Note that until now we have chosen these operators  from the Cartan
generators only ignoring their other choices 
 like
\be
g^+_j=e^{i\th^+ \sigma^+_{j}\tau^-_{j}}, g^-_j=e^{i\th^- \sigma^-_{j}\tau^+_{j}} 
\ll{Gtgpm}\ee
which we may
 use now in  transformation (\re{Gt}) either individually or 
in   combinations. Without detailed analysis of individual cases we
demonstrate here only a sample case of  
 $g^+$ on a noninteracting two-chain $XXX$ model:
$H^{0xxx}= \sum_j H^{0xxx(2)}_{jj+1}= -  \sum_j \vec \sigma_j \vec \sigma_{j+1}+  
\vec \tau_j \vec \tau_{j+1}. $ Using only the CR (\re{cr}) it can be easily
shown that this
transformation introduces  various  interacting terms including 
interchain hopping as well as pair production terms 
in the form:
\bea H^{xxx(2)}&=& \sum_j g^+_jg^+_{j+1} ( H^{0xxx(2)}_{jj+1})
(g^+_jg^+_{j+1})^{-1} \nonumber \\
&=& H^{0xxx(2)}-\sum_j[i\th^+(( 
\sigma^+_{j}\tau^-_{j}(\tau^3_{j+1}-
 \sigma^3_{j+1} )
+ \sigma^+_{j+1}\tau^-_{j+1}(\tau^3_{j}- \sigma^3_{j}))
-( \sigma^+_{j}\tau^-_{j+1}(\tau^3_{j}- \sigma^3_{j+1})
\nonumber \\
&+&
 \sigma^+_{j+1}\tau^-_{j}(\tau^3_{j+1}- \sigma^3_{j}) ) +
2(\th^+)^2 \sigma^+_{j}\tau^-_{j}
 \sigma^+_{j+1}\tau^-_{j+1}]
\ll{Hpmm} \eea
We can apply similar  transformation involving $g^-_a$ on top of the
Hamiltonian (\re{Hpmm}) to restore its hermiticity and generate additional
interacting term.
Another simpler choice of such gauge transformation,  given by
$
g^\pm_j(\sigma)=e^{i\rho^\pm \sigma^\pm_{j}}
$ and similarly for $\vec \tau$-operators,
 act like some rotation on the spin operators. Therefore the effect of such
transformations on  each anisotropic $XXZ$ spin chain  yields
\be H^{xxz(2)}
= H^{0xxz(2)}-(1-\Delta)( i\rho^+( \sigma^+_{j}\sigma^3_{j+1}+
\sigma^3_{j}\sigma^+_{j+1} )
+2(\rho^+)^2( \sigma^+_{j}\sigma^+_{j+1}
 +(\vec \sigma \to \vec \tau)).
\ll{Hpmma} \ee
It is clear that at $\Delta =1$, i.e. for the $XXX$ chain the additional
interactions in (\re{Hpmma}) vanish due the the rotational invariance of
the model.
 Note that in spite of the  additional interactions appearing in the models
like (\re{Hpmm}), (\re{Hpmma})
they remain quantum integrable by construction and 
 the associated Lax operators and
R-matrices can be constructed through 
 (\re{asg}) using  (\re{Gt}).
Note that though such transformations of the spin models were known in one
 or in
the other forms, their links with the quantum integrable systems and the
solutions of the YBE were perhaps never detected earlier. 
 However such gauge transformations, as we have
mentioned, do not affect the eigenvalue solution and
therefore the Bethe equations are given in the same form  as in 
   the original model.
\subsection{ New suppersymmetric t-J models with transformed interactions } 
We can use the freedom of generating operators (\re{G}) to construct
transformed fermionic, bosonic or spin models using the higher rank groups.
Here we shall consider such an application
using  $gl(1,2)$
 for constructing integrable variants of suppersymmetric $t-J$ model with
correlated hopping and other interactions along with their
 explicit Lax operator solutions. As we know
the 1d supersymmetric $t-J$ model with the constraint on the coupling
constants: $J=2t $ turns out to be an exact quantum integrable system with
the $R$-matrix and the Lax operator given as
\be \check R_{aj}(\la)=c(\la) I+ b(\la)\ti P_{aj}, \ \
 L_{aj}^{(tJ)}(\la)=b(\la) I +c(\la) \ti P_{aj}, \ll{RL} \ee
 where $b(\la)= {\la \ov \la+i}, \ c(\la)= {i \ov \la+i}$ and 
 $\ti P_{aj}$ is the graded  permutation operator formed by the generators
of 
 $gl(1,2)$ \c{koress}.

The operators (\re{RL})  satisfy now  a graded version of the YBE
 and the regularity condition
and therefore following the arguments of 
(\re{H})
the Hamiltonian of this SUSY $t-J$ model takes the form 
 $ H=\sum_j \ti P_{jj+1}$ .
Since we are interested in the   integrable extension of this t-J model
 with
correlated hopping, we can apply   
 (\re{GT}) on its Lax operator for generating  new set of 
 Lax operator and R-matrix 
 solutions of the 
YBE. We may construct the  
 transforming operators $S,T,G$ in the same form (\re{SG}), (\re{T}) and
(\re{Gt}) by using pair of operators from the set $n_+, n_-,n_0$. However due
to the constraint $n_++ n_-+n_0=I$ we are left with the only 
 choice of $n_\pm$
 yielding the transformed $t-J$ model as
\bea
\ti H&=&  \sum_{j} \left(-t{\cal P}\sum_{\sigma=\pm}
 c^\dag_{j+1 (\sigma)}c_{j (\sigma)} e^{-2i \gamma_{(\sigma)}
}  +cc.\right){\cal P}
\nonumber \\&+&  2t\left(\ha { S^+}_j{ S^-}_{j+1} k_{jj+1} +
 cc + { S^3}_j{ S^3}_{j+1}
  -{1 \over 4} {n}_j{n}_{j+1}\right)
  + {n}_j+{n}_{j+1}
 \ll{Htj}\eea
where ${\cal P}$ projects out double occupancies, ${\bf S}_j$
is the spin operator  and $n_j = n_{j(+)} +
n_{j(-)}$  the total number of electrons on site $j$.
Note that in (\re{Htj})  the operator parts in the correlated hopping 
 given by the  same form as (\re{S}) involving interactions
between $\uparrow$ and $\downarrow$ component electrons  vanish due to the  
constraint on double-occupancy indicated by the projector ${\cal P} $.
 However the operator parts  arise as  additional interactions in
the spin terms  given by 
\be
 k_{jj+1}=
e^{-i \left(2(\gamma(+)-\gamma(-)) 
+\sum_{\sigma=\pm}( \th  ( n_{j(\sigma)}+ n_{j+1(\sigma)})+(s-g)
 \sigma (n_{j(\sigma)}- n_{j+1(\sigma)})\right)}.
\ll{Sij}\ee
On top of this   transformation another one may be considered
 in the factorised form
(\re{Gt})  taking the factors as 
$
g^\pm_j=e^{i\rho^\pm S^\pm_{j}},$ which 
 is equivalent to a $SU(2)$ rotation.
  Using the commutation relations like 
$$ [S^-, Q_{-}]= [S^-, Q^{\dagger}_{+}]=0, \ \  [S^-, Q^\dagger_{-}]= 
 Q^{\dagger}_{+}, \ [S^-, Q_{+}]= - 
 Q_{-}, $$
where $Q_{\pm}=(1-n_{j(\mp)})c_{j (\pm)}, $ one can show that 
under such rotations, for example with $g^-_j$, the twisted hopping part of the 
SUSY $t-J$   model (\re{Htj}) would be  transformed to
\be
-t {\cal P}\left(\sum_{j,\sigma=\pm} 
 c^\dag_{j+1 (\sigma)}c_{j (\sigma)} e^{-2i \gamma_\sigma } +c.c. +
i\rho^-( c^\dag_{j+1 (+)}c_{j (-)})(e^{-2i\gamma_-} - e^{-2i\gamma_+})
 \right){\cal P} 
\ll{tjp}\ee
introducing hopping  between up and down electrons. Since 
$[S^-, n_{(+)}+n_{(-)}]=0, [S^-, n_{(+)}-n_{(-)}] \neq 0$, 
such transformations preserve the total charge but not the spin, 
though they have  no effect on the eigenvalue problem as we
have mentioned earlier. Nevertheless such  $t-J$ models as well as
(\re{Htj}) with additional 
interactions represent  integrable systems  associated with
 Lax operator  and $R$-matrices obtainable from (\re{GT}) starting with 
(\re{RL}). 

 We demonstrate  below  the application of the ABA method 
for solving the transformed models on the explicit examples of the 
Hubbard type 
 as well as the SUSY $t-J$ models  with correlated hopping.

\section {Algebraic Bethe ansatz for correlated hopping models}
\setcounter{equation}{0}
In a quantum integrable system the
 ABA method, which solves the EVP
 for the entire set of conserved
operators including the Hamiltonian 
by solving the EVP  for the transfer matrix,
 works 
also  with equal success for the transformed systems with correlated
hopping as introduced above. However, as
we   see  below,  among all the transformations discussed above 
only  the twisting transformations can influence
 the EVP and yield
 deformed Bethe equations.
Since the  applications of the ABA to the twisted model goes almost
 parallelly to
that for the original models,  
we present here only  the main steps of the ABA formulation for  the 
 twisted Hubbard (\re{Hubt}) and
the t-J model (\re{Htj}).

\subsection{ABA for the extended Hubbard model}
The Lax operator and the  $R$-matrix  
 for the 1d Hubbard model, which may be given by  two coupled  free
fermionic $R$-matrices
 of 6-vertex type, were known for quite some time \c{hubbardi}:
\bea
L^{(hub)}_{aj}(\la_a)&=&(L^{(+)}_{aj}(\la_a) \otimes 
(L^{(-)}_{aj}(\la_a)) exp(h_a
\sigma^3_{+ a}\sigma^3_{- a})\nonumber \\
R^{(hub)}_{ab}(\la_a, \la_b)&=&(L^{(+)}_{ab}(\la_{ab}) \otimes 
(L^{(-)}_{ab}(\la_{ab})) \cosh h_{ab} \cos \tilde \la_{ab}
\nonumber \\
&+&
(L^{(+)}_{ab}(\ti \la_{ab}) \otimes 
(L^{(-)}_{ab}(\ti \la_{ab}))
(\sigma^3_{+ a}\sigma^3_{- a})\sinh h_{ab} \cos  \la_{ab},
\ll{RLhubbard}\eea
where $L^{(+)} $ and $L^{(-)}$ correspond to
 $6$-vertex free fermionic models and involve  two independent
spin-$\ha$ operators $\sigma_{(+)}$ and $\sigma_{(-)}$
respectively, representing    two spin-components of the electron.
In  (\re{RLhubbard}) the notations $h_{ab}=h_{a}-h_{b},$ with $\sinh 2h_a=
{1\ov 4} U\sin 2\la_a$ and $  
 \la_{ab}=
 \la_{a}-
 \la_{b}, \ti 
 \la_{ab}=
 \la_{a}+
 \la_{a} $ denoting the dependence on the difference and the sum of the
spectral parameters have been used. However, in spite of this fact the 
explicit  ABA
formulation of the Hubbard model was discovered only
 recently \c{ramos}. This happened possibly 
 due to somewhat 
unusual tensorial structure  of its $16 \times 16$ $ R$-matrix  
and the $4 \times 4$ Lax operator. Consequently 
 the monodromy matrix has the form 
\begin {equation}
T(\lambda) = \left( \begin{array}{c}
 B(\lambda) \ \vec B(\lambda) \  F(\lambda) \\
    \vec C(\lambda) \  \hat A(\lambda) \
 \vec B^*(\lambda) \\
     C(\lambda) \ \vec C^* (\lambda)  \ D (\lambda)
   \end{array}   \right),
\ll{monod}\ee
 where $\vec B(\lambda)(\vec B^*(\lambda))$ and $ \vec
C(\lambda) (\vec C^*(\lambda))$ are two component vectors representing
one-particle creation and anihilation operators, while $\hat A(\lambda)$ is
 a $2 \times 2$ matrix with $A_{12}, A_{21}$ are the corresponding
spin excitation 
operators. The scalar operators $F(\la)$ and $C(\la)$ on the other hand 
correspond to two-particle  creation and annihilation  respectively.  
 The integrability condition here is given by a graded version of the YBE
(\re{ybeg}). The ABA becomes tricky due to the presence of a number of
creation/annihilation operators, which should be treated properly for
solving the eigenvalue problem of the transfer matrix
 $\tau(\la)=sTr [T(\la)], $
 generated by the supertrace of the monodromy matrix (\re{monod}):
\begin {equation}
[ B(\lambda) -\sum_{a=1}^2  A_{aa}(\lambda) \
 + D (\lambda)] 
| \la_1, \ldots , \la_N; \mu_1,\ldots , \mu_{N_\d} >= \Lambda_{N,N_\d}
(\la, \{\la_j\},\{\mu_\bet \})
|\la_1, \ldots , \la_N; \mu_1,\ldots , \mu_{N_\d}>.
\ll{eigen}\ee
This   in turn through expansion of $ \ln \Lambda_{N,N_\d} (\la)$ in powers
of $\la$, as seen from 
(\re{cn}), solves the eigenvalue problem simultaneously for all conserved
operators including the Hamiltonian of the system, e.g. the energy 
of the model should be given by $E_N=
 \Lambda_{N,N_\d}^{'} (0) \Lambda_{N,N_\d}^{-1} (0) $. The most important
point in solving the eigenvalue problem is to identify a
 $4 \times 4$ $R$-matrix of the $XXX$ spin
$\ha$-chain hidden  inside the  $R^{Hub}$-matrix of the Hubbard model,
which turns out to be  the two-particle scattering matrix
 $S$ of this system constituted out of the matrix elements of $R^{Hub}$
(\re{RLhubbard}):  
$
R_{6,6}=R_{11,11}, R_{7,7}=R_{10,10}, R_{10,7} = R_{7,10} $.
We have to adopt here the nested or repeated Bethe ansatz approach,
choosing the ferromagnetic vaccuum state as the  
pseudovacuum $|0>$ of the system, which gives 
the eigenvalue expression as  
\bea
 \Lambda_{N,N_\d}
(\la, \{\la_j\},\{\mu_\bet \})&=&
<0|B(\la)|0> \prod^N_j \left({R_{1,1} \ov R_{2,5}}
 (\la,\la _j) \right) + <0| D(\la)|0> 
\prod^N_j \left({R_{4,13} \ov R_{8,14}} (\la,\la _j) \right) \nonumber \\& -
&
\prod^N_j \left({R_{6,6} \ov R_{2,6}} (\la,\la _j) \right) 
 \Lambda_{N,N_\d}^{(1)}
(\la, \{\la_j\},\{\mu_\bet \}),
\ll{lambda}\eea
where the factor $\Lambda_{N,N_\d}^{(1)}$ related to the 
 second  step in the nested ABA is given by
\bea
\Lambda_{N,{N_\d}}^{(1)}(\la, \{\la_j\}, \{\mu_\bet \})&=&<0^{(1)}|
A_{11}(\la-\la_j)|0^{(1)}> 
\prod^{N_\d}_\bet \left({R_{6,6} \ov R_{7,10}}
 (\la,\mu _\bet ) \right)
\nonumber \\
 &+& <0^{(1)}| A_{22}(\la-\la_j)|0^{(1)}> 
\prod^{N_\d}_\bet \left({R_{11,11} \ov R_{7,10}} (\la,\mu _\bet ) \right), 
\ll{lambda1}\eea
where $|0^{(1)}>$ is the nested pseudovacuum with respect to the 
spin excitations.  
For identifying the   effect of  the gauge and twisting
transfornations we note that we will be concerned only on the 
 modifications of the  diagonal elements of the monodromy matrix  $T(\la)$
and 
  the elements of the scattering matrix $S$ involved in 
(\re{eigen})  through 
(\re{lambda}) and
(\re{lambda1}). 
It is however not difficult to check  through such explicit expressions  
 that the gauge transformations $G,S$ do not
 enter in any of these elements and therefore do not have any effect on the 
eigenvalue formulas and consequently on the   Bethe ansatz equations.
 This is the reason why the Zvyagin et al 
and the {\it minimal} Schulz-Shastry  models related
 through a gauge transformation  exhibit
 same Bethe equations as explained above. 
We find on the other hand 
 that under the twisting transformation (\re{T}) the
relevant terms that suffer changes  are the transfer matrix operators
 $A_{22}(\la-\la_j)|0>^{1}=e^{-i2\th N }\prod_j^N b(\la-\la_j)
 |0^{(1)}>$  
   and the element of the scattering matrix
 $R_{7,10}(\la)=b(\la) e^{-2i
\th}, $  where
$b(\la)={\la \ov \la+U}$. For simplicity we shall consider only the parametr 
$\th$ in the twisting transformation putting $\gamma_({\pm})=0$.
   Incorporating these changes in 
(\re{lambda1}) we get its 
modified equation under twisting as
\be
\Lambda_{N,{N_\d}}^{(1)}(\la, \{\la_j\}, \{\mu_\al\})= e^{2i\th {N_\d}}
\prod^{N_\d}_\bet {1 \ov b(\mu_\bet-\la )} +  
 e^{2i\th ({N_\d}-N)} \prod^N_j b(\la -\la_j )
\prod^{N_\d}_\bet {1 \ov b(\la -\mu_\bet )}.\ll{lambda11}\ee
Notice that  at $\la=\la_j $ and $ \mu_\bet $ singularities appear in some of the 
terms in the expressions for the eigenvalues (\re{lambda}),(\re{lambda11}).
Therefore demanding the vanishing of residues at this points
we can  arrive  at the Bethe equations for determining 
the parameters $\{\la_j\}, \{\mu_\bet
\} $.
The residue of $ \Lambda_{N,N_\d}(\la)$ at $\la=\la_j$ yields
\be
<0|B(\la_j)|0> \equiv e^{i p_j L}=
e^{i2\th N_\d} \prod_{\al=1}^{N_\d } { \mu_\al -\la_j +{U }
\over \mu_\al -\la_j },
\ll{betheeqn1} \ee
   while the same at $\la=\mu_\bet $
gives
\be 
e^{i2\th N} \prod_{j=1}^N { \mu_\al -\la_j +U
\over \mu_\al -\la_j}
=\prod_{\bet=1}^{N_\d} { \mu_\al -\mu_\bet +{U}
\over \mu_\al -\mu_\bet -{U }},
\ll{betheeqn2} \ee
 representing  the Bethe equations for the integrable 
 Hubbard model with correlated hopping.
Note that for $U \to {i \ov 2} U$ the Bethe equations (\re{betheeqn1})
and (\re{betheeqn1}) coinsides exactly with the CBA results corresponding to
the Hubbard model with correlated hopping considered in
\c{zvyag92,SS98,kun98}.
Note again that  the system of \c{borov912,zvyag92}
 is gauge
euivalent to   the {\it  minimal} model of \c{SS98},
 which coincides also with that of \c{kun98}. We remind that this
equivalence holds at the Lax operator and the application of the ABA level,
which does not require explicit form of the Hamiltonian. The equivalence
however extends also   
to their Hamiltonians, provided they are 
  crrectly derived from the Lax operators.

\subsection{ABA of twisted $t-J$ model}
The nested ABA treatment  for the twisted Hubbard model presented above,
 may be carried out in a similar way for the twisted  $t-J$ model.
Interestingly analogous to the Hubbard model, we also find here that, though the
 twisting transformation $T$
as well as the 
 gauge transformations $G,S$ considered above contribute in generating the
 correlated hopping terms in the Hamiltonian, only the former type of
 transformation affect the ABA treatment and the related Bethe equations.
 The essential ingredients needed for the ABA application to this system,
i.e.
 the transformed  Lax operator and the $R$-matrix  are  given as in 
(\re{LLt})
with
(\re{RL}) leading to the corresponding monodromy matrix in the form
\begin {equation}
T(\lambda) = \left( \begin{array}{c}
A_{11} (\lambda) \ A_{12} (\lambda) \  B_1(\lambda) \\
    A_{21} (\lambda) \   A_{22} (\lambda) \
  B_2(\lambda) \\
     C_1(\lambda) \  C_2(\lambda)  \ D (\lambda)
   \end{array}   \right).
\ll{monodtj}\ee
 Since the ABA steps for this twisted model is close to those for the
original supersymmetric $t-J$ model, we will follow the formalism of
\c{koress} pointing out only the essential differences under twisting. This
supersymmetric model involves the bosonic (B) hole state together with the
fermionic (F) electron states and we  consider the $BFF$ case for
concreteness, where
 in the first step of nesting $|0>$  represent 
the  hole state, over which
 $C_1,C_2 (B_1,B_2)$ act as  the creation (annihilation) operators  producing
fermionic spin-$\d$ and $\up$ pseudo-particles with charge excitations.
  Spin excitations appear in the next step with $A_{21} (A_{12})$ becoming
the creation (annihilation) operators over the vacuum $|0^{(1)}>$ mimicing the 
$XXX$ model. 
The conserved
quantities are again  given by the
supertrace \be sTr [T(\la)]\equiv \tau(\la)=-( 
 A_{11}(\lambda)+A_{22}(\lambda))+ 
  D (\lambda).\ll{evaluetj}\ee
The eigenfunctions are constructed as 
\be |\{\la^{(0)}_j \}_1^{M_0}, \{\la^{(1)}_\bet \}_1^{M_1}>= 
 \prod^{M_0}_jC_{a_j}(\la^{(0)}_j)|0>F^{(1)}_{a_1\ldots
a_{M_0}}(\la^{(1)}_\bet ),\ll{evectj}\ee where  
$F^{(1)}_{a_1\ldots  a_{M_0}}(\{\la^{(1)}_\bet\} )=\prod^{M_1}_\bet 
A_{12}(\la^{(1)}_{\bet} |0>^{(1)} F^{\bf a}_{\bf b} .$ Therefore for finding out 
the eigenvalue we have to consider carefully  the  commutation relations
between the elemets of $\tau(\la)$  (\re{evaluetj})  and the creation operators
forming the eigenvector (\re{evectj}) using the matrix relations 
of the 
graded YBE. This yields finally the eigenvalue expression
\be
 \Lambda_{M_0,M_1}
(\la, \{\la^{(0)}_j\},\{\la^{(1)}_\bet \})=
b(\la))^L\prod^{M_0}_j  {1 \ov b(\la-\la_j^{(0)} )}\Lambda^{(1)}_{M_0,M_1}
(\la) +\prod^{M_0}_j  {1 \ov b(\la_j^{(0)} -\la)}
\ll{lambdatj}\ee
where 
\be
\Lambda^{(1)}_{M_0,M_1}
(\la)
=-\left (\prod^{M_0}_j  { b_+(\la-\la_j^{(0)} )}\prod^{M_1}_\bet
  {1 \ov b_-(\la_\bet ^{(1)} -\la)}+ 
\prod^{M_0}_j  \left ({b(\la -\la_j^{(0)}) \ov b(\la_j^{(0)} -\la)}
\right)
\prod^{M_1}_\bet
  {1 \ov b_-(\la -\la_\bet ^{(1)})}\right)
\ll{lambda1tj}\ee
$$$$$$$$
Note that the effect of twisting transformation in the above eigenvalue
 expressions 
can be detected  in the terms
$b_\pm(\la)= {\la \ov \la+i}e^{\pm2i \th},$ where the term $b_+$ 
 comes from the transfer matrix element, while   $b_-$ is the $R$-matrix
element $R_{2,5}$ appearing also in the two-particle scattering matrix.
The Bethe equations ,
 as discussed above, can be derived from the analiticity condition of
 the egenvalues yielding 
\bea  (b(\la_j^{(0)})^{-L}
 \equiv e^{i p_j L}&=&
e^{i2\th M_1} \prod_{\bet=1}^{M_1} {\la_j^{(0)}  -\la^{(1)}_\bet +{i}
\over \la_j^{(0)}  -\la^{(1)}_\bet  },
\nonumber \\
e^{i2\th M_0} \prod_{k=1}^{M_0} {\la^{(1)}_\al -\la_k^{(0)}   
\over \la^{(1)}_\al -\la_k^{(0)}-i }
&=&\prod_{\bet=1}^{M_1} { \la^{(1)}_\al -\la_\bet^{(1)}+i
\over \la^{(1)}_\al -\la_\bet^{(1)} -i},
\ll{be2tj} \eea
Note the similarity between the   Bethe equations  for the twisted
$ t-J$ and  the Hubbard model. However they do differ due to the different
structure of their $R$-matrices and the choice of vacua.
The CBA analysis \c{sarkar}  of the twisted  t-J model (\re{Htj})
  is expected  to lead also to
the same Bethe equations as above, which  appeared also in \c{zvyag92}.   
 
\section{Concluding Remarks}
\setcounter{equation}{0}
We have identified a class transformations including gauge and twisting
transformations by exploiting the symmetries of the Yang-Baxter equation, 
 which generates multi-chain integrable systems
of correlated hopping electron  or spin models with interachain interactions. 
Comparing 
our construction with the classification
 of such CBA solvable models made in \c{oster00}, we 
  identify an  important
 quantum integrable subclass within this CBA solvable models.
The solvable models of \c{oster00}, which are not covered by our
scheme might still show integrability, though possibly they would
fall under nonultralocal models \c{kunhlav} not considered here and need seperate
investigation. 
 On the other hand  our construction can go
beyond the  models classified in \c{oster00}
 and generate other integrable extentions of the  $t-J$ and spin
models as we have demonstrated here.
 We  have applied the ABA method  to  the  Hubbard and the SUSY $t-J$ models 
with correlated 
hopping, which are integrable 
models constructed  through our scheme.
The integrable models allow mutually commuting higher conserved
operators and the exact solutions of 
 their eigenvalue problem through the  ABA treatment. Our findings 
shows explicitly that only the twisting transformations need to  be
incorporated  in the ABA equations, while the gauge transformations 
 have no effect on them, though both of these transformations 
 can considerably change the  form of the Hamiltonian. Therefore
this draws the important conclusion that  all
models related by   gauge transformations share the same ABA equations 
and the eigenvalues, which is in fact the case with the Zvyagin et al model
and the {\it minimal} model of Schulz and Shastry.
Since our method is based on the transformation of the Lax operator
associated with the model and not of the Hamiltonian 
as done in \c{oster00}, we can get more general information about such
transformations.
  Starting from this
general scheme we are  not only able to  find the explicit 
relationship between different models
known in the literature at their Lax operator and Hamiltonian level,
but also could detect some crucial errors in the derivation of a
well known  model, which
resolves  completely  the recent
controvercies  around the equivalence and solvability 
of   some known  models.

Application of the present scheme of constructing twisted and gauge
transformed quantum integrable systems to other models of physical
interest, like spin ladder \c{spinlad}, $t-J$ ladder \c{tjlad}, Bariev model
\c{bariev} should be interesting problems to carry out.
Recall also that the transport property in strongly correlated electron
systems  is probed through  the Drude weight \c{drude}, which  
is usually calculated using the twisted boundary condition caused by the 
 parameter $\gamma_{(+)}=\pm\gamma_{(-)}=\phi_{\pm}$
 in the twisting transformation 
like (\re{T}). Therefore the physical relevance of the other 
twisting parameter
$\th$ considered here, as well as its possible effect in extending the notion 
of the Drude weight should be explored \c{prep}.

It should be mentioned also that various applications and generalizations of 
the  twisting transformation 
\c{twist} can  be found   in \c{twist9498,twist9801} 
and the references therein.  
  The most  relevant among them in
the present context is \c{twist9801}, where twisting transformations
 were applied to the
supersymmetric $t-J$ and $U$ models for deriving their Hamiltonians as well
as different forms of the
Bethe ansatz equations  modified by such twisting.
 
\medskip

\ni{\it Acknowledgment}: The author thanks the AvH Foundation of Germany
for its support through its Followup Programme and the anonymous 
referee for pointing
out some recent references. 

\end{document}